Structures in invariant mass distributions from (μ,π) combinations in the range $0.380 < M_{\mu\pi} < 0.470$ GeV from neutrino and kaon experiments are presented in comparable formats. No artifacts have been found to account for any part of the structure. Hypotheses that the similarities are due to recurrent statistical fluctuations are beyond credibility. My conclusion is that the similarities are overwhelming evidence that the structure is of an unexplained physical origin. It includes an enhancement which would accord with the decay of a narrow (μ,π) state of mass 0.429 GeV. The purpose of this report is to request and enable every experimenter with precise $M_{\mu\pi}$ distributions to investigate their degree of correspondence with these analyses.
\\

 **Introduction:**   This report summarises evidence of physical structure in (μ,π) invariant mass ($M_{\mu\pi}$) distributions in the range $0.380 < M_{\mu\pi} < 0.470$ GeV, from neutral combinations produced in some 11 independent neutrino and kaon experiments. No tenable hypothesis of artifacts, or of a concurrence of random fluctuations, has been found to account for any of the structures.  The only logical explanation that can be proposed for the detailed similarity in independent experiments is that it is of a physical origin.  The most significant enhancement in the structures accords with an hypothesis that it is due to the decay of a short lived (μ,π) state of mass 0.429 GeV.

   Adequate information for a meaningful understanding of these structures, particularly for the identification and kinematics of the assigned "(μ,π)" over all configurations, is unlikely to be acquired without specially designed experiments. Nevertheless, there are the precedents of significant contributions to the present studies from carefully executed kaon and neutrino experiments, designed without any initial intent to investigate structure in $M_{\mu\pi}$ distributions. My present purpose is to encourage every group of experimenters with information about $M_{\mu\pi}$ distributions in the range $0.380 < M_{\mu\pi} < 0.470$ GeV to examine the extent to which their own observations correspond with these analyses, and to consider developing experiments to discover causes of the structure.  For that reason the information here is in a format to enable readers with such observations to compare them directly with these distributions.

   Readers not concerned with such comparisons may wish to shorten the text by omitting the notes, Table 1 and the appendices.

(1) **References and diagrams**

   The sections and diagrams indicated by (II) are those in  my 1985 report[1] which will serve here as a catalogue of information.  In (II), and here, (I) refers to an earlier publication[2]. (A) indicates appendices concerning details which may be of use to readers not usually concerned with (μ,π) combinations.   For the comparison of $M_{\mu\pi}$ distributions, without distraction by mean levels of differing slopes, curvatures and backgrounds, most diagrams will be in a format similar to Fig.6 in (II) where the histograms are plots of the excursions of the bin contents from a polynomial mean[3]. In those diagrams, the scale of the ordinate was chosen to make one standard deviation the same length for all. Here, to facilitate comparisons of structure in the distributions, independently of the unknown level of a background continuum, the mean passes through the distribution at 1.0 and the scale distance[4] between the extreme upward and downward departures from the mean is the same in each figure. Also to facilitate comparisons, the same diagram is sometimes shown in more than one figure. Information about the data sources is in Table 1. Notes and references are as usual.

(2) **Evolution of these studies**

At the time of the neutrino and kaon experiments in the CERN heavy liquid bubble chamber (HLBC) it was generally believed that all $M_{\mu\pi}$ distributions would be statistically smooth. The observation of an unexpected accumulation of values in the range $0.415 < M_{\mu\pi} < 0.435$ GeV from interactions of the type $\nu_\mu + N \rightarrow \mu^- + \pi^+ + N...$ did not seem to accord with the belief. In and near that range of $M_{\mu\pi}$ more numerous data were available from entirely independent experiments with $K^0_{\mu 3}$ decays in the HLBC and in the Brookhaven 14 inch hydrogen bubble chamber (HBC). Neither of those $M_{\mu\pi}$ distributions supported a postulate that they were statistically smooth: there were unexpectedly similar structures in the vicinity of an enhancement determined, collectively, to be at $0.429 \pm 0.002$ GeV. As shown and referenced in (II), $M_{\mu\pi}$ distributions kindly made available from spark chamber experiments with kaons accorded with the bubble chamber observations. Statistical estimates set a probability of $<10^{-10}$ that the enhancements in the "0.429 bins" were all due to random fluctuations. (The bin containing 0.429 Gev will be called the "0.429 bin").

As will be examined in more detail in section 5, in a neutrino experiment with average event energies of 90 GeV and a mass scale "that cannot be in error by >0.003 GeV" Ballagh et al.[10] reported an "enhancement" in their "0.429 bin", which they "believe is a statistical fluctuation". Allasia et al.[17] have reported that in their $M_{\mu\pi}$ distributions with average event energies of 45 and 62 GeV "no significant $M_{\mu\pi}$ enhancement is observed". Neither Ballagh et al. nor Allasia et al. indicated their reasons for not showing a scientific comparison of their distributions with other observations. Comparisons in this note with $M_{\mu\pi}$ distributions in (II) from earlier neutrino experiments, make clear that those groups of authors were unaware that they had obtained, at higher energies, informative confirmations of the earlier observations.

(3) **Experimental limitations with (μ,π) combinations**

In only some 2% of the (μ,π) combinations here is there experimental identification of either of the assigned μ or π. For the rest, assignment in each combination is by some convention. In (μ,π) from neutrino experiments, in the absence of specific evidence, the assignment of muon is given to the particle with charge appropriate for the neutrino or antineutrino beam. Since all such beams contain a component of the opposite lepton number, there are incorrect assignments. In experiments with kaon decays, because there has been no practical means of identification of pions and muons over a wide range of momentum, a first consideration is the selection of neutral combinations likely to be (μ,π). Their assignment may then be according to the laboratory momenta of the particles (e.g. Fig.2(a)), or simply by including both possible calculated $M_{\mu\pi}$ values in the distribution (e.g. Fig.2(b)). The effects of incorrect assigment must be considered for each $M_{\mu\pi}$ distribution (A1).

It will be seen that in two experiments with spark chamber detectors, designed to investigate other aspects of kaon decays, their particular constraints contributed valuable information (Figs.3(a) and 8(d)) to these studies although, initially, the observations were not understood. While detailed angular distributions of the $(\mu,\pi)$ which contribute to the structures remain unknown, constraints on their selection can continue to produce unexpected results, e.g., a constraint on the lower limit of detectable momentum can reduce the 0.429 GeV enhancement.

As in (II), most of the diagrams in this note refer to $0.380 < M_{\mu\pi} < 0.470$ GeV. This range has been used for these studies because it is symmetrical about the "original" enhancement, and it does not extend to known[5] artifacts near each end (A2). Details of various data sets are in (II), summarised here in Table 1. All mass scales were calibrated with reference to the kaon decay to $(\pi^+,\pi^-)$ (A3).

(4) **Features observed at and near $M_{\mu\pi}$ = 0.429 GeV**

(a) **An enhancement in the "0.429 bin":** From the first studies it was evident that the most significant and constantly located feature in $M_{\mu\pi}$ distributions in the selected range was the enhancement at 0.429 GeV.

Despite the searches for contamination of the "$(\mu,\pi)$" by other particles which might cause this particular feature (A2), and the fact that for only very few combinations is there any direct identification of the muon or pion, no reason has been found to doubt that the enhancement at 0.429 GeV is caused by $(\mu,\pi)$ combinations. In the continuing absence of any other hypothesis, it was postulated[6] that it is due to a narrow $(\mu,\pi)$ state with a decay $M_{\mu\pi}^* \rightarrow \mu^- + \pi^+$ and, with opposite charges, also for the antiparticle. The mass of such an $M_{\mu\pi}^*$ is 0.429 ± 0.002 GeV.

Enhancements in the "0.429 bin" are shown in Fig.1: (a) is from the first neutrino experiment in the HLBC, (b) is from the combined neutrino data from the HLBC and from Gargamelle (average event energies <10 GeV), and (c) is from neutrino interactions at average energies of 90 GeV in the Fermilab 15-ft hydrogen bubble chamber. Similar distributions from bubble chamber experiments to study $K^0_{\mu3}$ decays are in Fig.2: (a) is from the HBC, (b) is from the HLBC and (c) is from the CERN 2-m hydrogen bubble chamber (all data are referenced in Table 1).

The invariant mass ($R_{\mu\pi}$) calculated from the reverse assignment of the muon and pion is equal to $M_{\mu\pi}$ only in those combinations in which the two particles have the same laboratory momenta (see (I), section IV and Fig.5). Thus in all $M_{\mu\pi}$ distributions with incorrect assignments, any enhancement due to a narrow state is associated with $R_{\mu\pi}$ from the incorrectly assigned $(\mu,\pi)$ which, otherwise, would have contributed to the enhancement. Their values depend on the energy of the parent $M_{\mu\pi}^*$ and its decay configuration. Next we consider other features observed in most of the $M_{\mu\pi}$ distributions near, but not at 0.429 GeV.

(b) **The "lobes":** In the data of Fig.1(b) and in Fig.2(a),(b),(c), and in spark chamber observations (e.g., in (II) Fig.6(e),(g) which are reproduced in the format of this note in Fig.8), the next most noticeable features near the $M_{\mu\pi}^*$ enhancement are seemingly associated enhancements, called "lobes" in (II). Their occurrence in so much of the data makes them as unlikely as the $M_{\mu\pi}^*$ enhancement to be random fluctuations. However, unlike the invariant position of $M_{\mu\pi}^*$, the mean position of either lobe differs slightly in the various data sets: (a), (b) and (c) in Fig.2 have been ordered according to the increasing separation of the lobes (which also corresponds with increasing average energy of the kaon beams).  That such an ordering can be made, while the $M_{\mu\pi}^*$ enhancement is in a constant location, makes it unlikely that the lobes are caused by similar nearby narrow ($\mu,\pi$) states. The $R_{\mu\pi}$ from $M_{\mu\pi}^*$ of different laboratory energies could only cause narrow lobes if the decay configuration were suitably non-isotropic with respect to its line of flight.

   Inspection of these various distributions leads to the question: why are there no significant indications of lobes near the enhancement in Fig.1(a)?  Information that seems relevant is that data set is the most likely of all the sets to have correctly identified pions[7].  The conjecture that the lobes are not due to ($\mu,\pi$) combinations, but to ($\mu,\mu$) combinations will be further examined.  (Sections VI et seq. of (II) venture an hypothesis for the origin of lobes.)

  (c) **"Depletions"** <u>occurring with the enhancement</u>: On each side of all the $M_{\mu\pi}^*$ enhancements there are regions where the data level is below the level of a polynomial mean (e.g. in Fig.1 (b),(c) and in Fig.2).  Even if the means are calculated for the histograms with the bins of the enhancement removed, the depletions are little changed.

   Angular distributions in the $M_{\mu\pi}^*$ centre of mass system and selection procedures for the ($\mu,\pi$) combinations can be postulated for which the $M_{\mu\pi}$ and $R_{\mu\pi}$ could cause apparent data depletions at each side of the principal enhancement.  This fact does not lead to a convincing explanation that the accumulation of $R_{\mu\pi}$ in other regions is the main cause of these depletions.  As mentioned in section 3, without more identified data with accurately known experimental conditions for their detection, the actual angular distributions in the production of the enhancement and the lobes are likely to remain unknown, together with the causes of the depletions adjacent to the $M_{\mu\pi}^*$ enhancement.

  (d) **"Depletions"** <u>instead of the enhancement and lobes</u>: Figs.4(d) and 17(a) in (II) show depletions in an $M_{\mu\pi}$ distribution[8] from kaon decays at locations where the "usual" 0.429 GeV enhancement and lobes would have been expected.

   Initially, this was an inexplicable result from a precision experiment, designed to study two body decays of $K^0_L$ with high statistics.  The $M_{\mu\pi}$ distribution is shown in Fig.3(a) in the format of the preceding diagrams: Figs.3(b) and (d) are the kaon observations of Figs.2(b) and (a), repeated for comparison.  (The average momentum of the kaon beam for Fig.2(b) is the nearer to that for Fig.3(a).)   There is a correspondence between the enhancement and lobes in Figs.3(b) and (d), and the three principal depletions in Fig.3(a).

   Those depletions can be made to simulate the appearance of the "usual enhancement and lobes" by inverting the diagram in the vertical sense to become Fig.3(c).  The apparent "0.429 Gev enhancement" and "lobes" in this diagram are not, of course, a measure of "missing enhancements" - they simply represent the data that would be required to fill the depletions in Fig.3(a) to the level of the mean.

   The criteria for selection of the combinations are the causes of the depletions in this $M_{\mu\pi}$ distribution: to reduce the background from other than two-body decays, the experiment was designed to record only combinations with equal laboratory momenta and no resultant momentum transverse to the line of flight of the parent kaon ($p_T=0$). Evidently, for all two-body decays $p_T=0$ and the phase space for isotropic decays is maximum for combinations with particles of equal laboratory momenta.  Some combinations from the "usual" $K^0_{\mu3}$ decays fulfil the selection criteria. The (μ,π) combinations in the "0.429" enhancement have, preferentially, $p_\mu \neq p_\pi$ and $p_T \neq 0$; still less of these combinations will be selected - hence the depletions. (In section 6, the result of the selection of combinations for which $p_T \sim p_{Tmax}$ will be discussed).

Fig.3(a) shows, as would be expected, that if in the $K^0_{\mu 3}$ decay a particle were formed which leaves no record in the detection system e.g., because of the kinematical selection criteria, (or because it is long lived), it may cause a depletion in the $M_{\mu\pi}$ distribution. The observation that the $M_{\mu\pi}^*$ enhancement and the usual lobes were together replaced by depletions accords with an hypothesis that the lobes are associated with $M_{\mu\pi}^*$.

(5) **Data from higher energy neutrino experiments**

(a) **H.C. Ballagh**[9] and 36 co-authors described an experiment on dimuon production by neutrinos in the Fermilab 15-ft bubble chamber, filled with a neon-hydrogen mixture and equipped with an external muon identifier (EMI). The average event energy was 90 GeV: the average event energy of previously investigated neutrino data was < 10 GeV. From the "326,000 good neutrino pictures with EMI information" there were 57 ($\mu^-,\pi^+$) and 14 ($\mu^+,\pi^-$) combinations with $M_{\mu\pi}$ in the range $0.380 < M_{\mu\pi} < 0.470$ GeV. Unlike earlier data, these were the first to be recorded in an experiment with an EMI.

H.C.Ballagh[10] and 26 co-authors, later, showed histograms of $M_{\mu\pi}$ from ($\mu^-,\pi^+$) and ($\mu^+,\pi^-$) combinations, in bins of 0.010 GeV, in the range $0.260 < M_{\mu\pi} < 1.200$ GeV (their Fig.2). They observed an enhancement in the range 0.420 to 0.430 GeV (i.e. in their "0.429 bin") which they "believe is a statistical fluctuation".

Their distributions of $M_{\mu\pi} < 0.520$ GeV, a greater range than in the preceding figures, have been reproduced in Fig.4(b) and (e), together with polynomial means[11] which have been fitted over the entire range of each data set in their Fig.2. The corresponding distribution for Fig.1(a), from the neutrino experiment of some 15 years earlier, separated into the ($\mu^-,\pi^+$) and ($\mu^+,\pi^-$), are shown in Fig.4(a) and (d). A mean is not needed for the inspection of those data (in any case there is one in the original report).

A first examination reveals similarities between Fig.4(a), (b), and (d), (e)[12] in the region of the "0.429" bins. It may be asked whether the $M_{\mu\pi}$ distribution of Ballagh et al. has other fluctuations similar to that in Fig.4(b) in the "0.429 bin"? Fig.4(c) shows the values of $(y-m)/(m)^{1/2}$ ( where y is bin content and m the level of the mean) for each of the 94 bins of the $M_{\mu\pi}$ distribution from the ($\mu^-,\pi^+$) in their Fig.2. Essentially, this diagram shows the number of standard deviations in the departure of each bin of the distribution from its polynomial mean. The maximum value of 3.1 standard deviations occurs in bin 17, which is their "0.429 bin": their leftmost bin is numbered 1.

Thus, if $(y-m)/(m)^{1/2}$ can be considered as a measure, Ballagh et al. observed the most significant of the fluctuations in the entire range of their histogram in 0.010 GeV bins of $M_{\mu\pi}$ from ($\mu^-,\pi^+$) combinations, in their " 0.429 bin". Since the probability is vanishingly small that random fluctuations cause all the previously observed enhancements in that bin, their "0.429 bin" could be described as the "expected bin" for an enhancement in their distribution. The probability of finding a 3.1 standard deviation in any bin is <0.004, the probability of finding the largest fluctuation in any specified one of their 94 bins is <0.02. Hence, according to the usual approach, the probability is $<10^{-4}$ for such a statistical fluctuation in the "expected bin".

It is contrary to usual statistical concepts to consider that random fluctuations cause reproducible patterns in distributions: the resemblances of the patterns in Fig.4(a), (b), (d) and (e) may be of more interest to some readers than an arithmetic of probabilities of bin content. Any structures can be better compared in Fig.5, in the same format as preceding diagrams, without the rising mean level of the data. Fig.5(a) shows the neutrino data in Fig.1(b) from the HLBC and Gargamelle, repeated for reference. Fig.5(b) is from the $(\mu^-,\pi^+)$ of Fig.4(b); (c) is from the $(\mu^+,\pi^-)$ data of Fig.4(e); (d) is from the total data[13] of Ballagh et al. in the range: it has been seen already as Fig.1(c). As shown in Table 1, Fig.5(a) refers to 368 combinations, (b) to 57 combinations, and (c) to 14 combinations (with the most content in any bin of 4).

In Fig.5, each distribution has its largest upward departure from the mean at, or adjacent to the "0.429 bin", each has noticeable "depletions". These features show that Ballagh et al. observed enhancements in $(\mu^-,\pi^+)$ and $(\mu^+,\pi^-)$ invariant mass distributions from a neutrino experiment with average event energies of 90 GeV, similar to those observed in neutrino experiments at event energies of some 10 GeV: that in itself is an important observation[14], even if expected.

What may be particularly significant is that features which could be described as "lobes" at the same masses as those in Fig.5(a), are not as pronounced (if they can be found at all) in Figs.5(b),(c) or (d). Because of the primary experimental aim to study dimuons it might be presumed that the "$(\mu,\pi)$ combinations" of Ballagh et al. are likely to contain fewer $(\mu,\mu)$ combinations than most of the data from other experiments. The authors gave no information in this regard: if the expectation is justifiable, then their observation would be important information for understanding the lobes.

In the enhancement in the reported total $M_{\mu\pi}$ distribution, Ballagh et al. observed some 8.5 events above the level of the mean value of, also, some 8.5 events. In terms of AE[15] in Table 1 their "0.429 bin" and its neighbourhood show a higher "signal to background" than all but the "original" enhancement. (For that observation it was appropriate to choose the binning to demonstrate the unexpected group of events). The absolute right of Ballagh and co-authors to interpret their observations in accord with their own wishes would be universally acknowledged. Some readers who compare what Ballagh et al. actually found, with previously reported observations, which the 27 coauthors did not, might not share the belief that the enhancement in their "0.429 bin" "is" a statistical fluctuation[6]. Of course, as for all of the others, it could be: in any case it does make still more unlikely any postulate that all the observed enhancements in the "0.429 bins" are statistical fluctuations.

(b) **D.Allasia**[16] and 47 co-authors described an experiment for the study of neutral strange particles in neutrino and antineutrino interactions in the BEBC bubble chamber, filled with deuterium and equipped also with an EMI. They reported 125,000 pictures from an antineutrino beam and 39,000 pictures from a neutrino beam. Average event energies were 45 GeV for antineutrinos and 62 GeV for neutrinos.

D.Allasia[17] and 36 co-authors, later, reported a search for enhancements in their $M_{\mu\pi}$ invariant mass distributions. They indicated that "(i) because of the selection criteria used to identify charged current events, all muons which enter in the mass combinations have momentum greater than 4 GeV/c and (ii), almost all charged kaons and all protons with momentum greater than 1.0 GeV/c are labeled as pions." In Fig.1 of their paper they show histograms in 10 MeV bins, from 232 ($\mu^-,\pi^+$) and 143 ($\mu^+,\pi^-$) combinations, in the range 260<$M_{\mu\pi}$<750 MeV. Their mass resolution "at around 500 MeV is about 9 MeV".

The constraint on the lowest muon momentum limits the observable range of $p_1/p_2$: this can reduce the $M_{\mu\pi}^*$ enhancement. Figs.6(a) and (b) show the distributions of $M_{\mu\pi}$<520 MeV for the ($\mu^+,\pi^-$) and ($\mu^-,\pi^+$) combinations of Allasia et al., (c) is the total of the data; each distribution with a polynomial mean. The ($\mu^+,\pi^-$) are chosen for a first inspection because protons are unlikely to be included as negative pions. Fig.6(d) is made from the total data similarly to Fig.4(c) for the ($\mu^-,\pi^+$) combinations of Ballagh et al. The values marked 1,2,3 in Fig.6(d) are for the bins so marked in Fig.6(c): 1 and 2 correspond to the bins in the region of where lobes might have been expected, 3 is in the region where the ($\pi^+,\pi^-$) artifact[12] can occur. It can be seen that the fluctuations where lobes might be expected are among the more significant upward fluctuations.

As in Fig.5, Fig.7(a) is again the $M_{\mu\pi}$ distribution of Fig.1(b) as a reference, (b) is from the ($\mu^+,\pi^-$) combinations of Fig.6(a), (c) is from the sum of the ($\mu^-,\pi^+$) and ($\mu^+,\pi^-$) combinations of Ballagh et al., (d) is from the ($\mu^-,\pi^+$) combinations in Fig.5(b). The more recognisable features in the distributions from Allasia et al. are the lobes near 0.40 and 0.45 GeV. They appear in each of the two regions in each distribution where they would have been expected. There is also some semblance of an upward fluctuation in or adjacent to the "0.429 bin". From an inspection of the distributions of Fig.6 it appears that without any scientific comparison with other observations, the conclusion of Allasia et al. that "no significant mass enhancement is observed" is all that they could report.

From Fig.7 it appears that the observations of Ballagh et al. and Allasia et al. differ in that there is no obvious evidence of lobes[18] in (c) in places similar to those in (a), (b) and (d). Allasia et al. do not state whether their "($\mu,\pi$)" combinations could also include ($\mu,\mu$), as conjectured in (II).

(6) **Spark Chambers**

So far this note has been mostly concerned with structure in $M_{\mu\pi}$ distributions from bubble chambers: it is important to make clear that corresponding structure is observed in spark chamber experiments. Figs.8(a) and (b) show structure in $M_{\mu\pi}$ distributions from different laboratories (see Figs.4(b) and (c) in (II)). The structure of Fig.2(b) from the HLBC is shown again as Fig.8(c) for comparison. Fig.3(a), from a spark chamber experiment in still another laboratory, is repeated as Fig.8(e) to recall the consequence of the selection $p_T$=0, already discussed in section 4(d).

When momentum is determined from track curvature in a magnetic field, knowledge of the charge of the particle is essential, but not of its mass: this fact permits the determination of the invariant mass of some neutral combinations without identification of the components. In the cms of the decay of either $K^0_{e3}$ or $K^0_{\mu3}$ the electron or muon neutrino momentum is $p_\nu=(m_K^2-m^2)/2m_K$, m is the invariant mass of either an (e,$\pi$) or ($\mu,\pi$) combination. Thus $p_\nu$ is a direct measure of the invariant mass of the combination: for $M_{\mu\pi}$=0.429 GeV, $p_\nu$=0.064 GeV/c. If both (e,$\pi$) and ($\mu,\pi$) combinations are present, as is inevitable, and the selection $p_T$=0 is unfavorable for the observation of the "usual" structure in the $M_{\mu\pi}$, as shown by Fig.3(a), then it might be asked whether the $p_T$ distribution in the vicinity of 0.064 GeV/c might show some evidence of the "usual" structure?

In that regard the suggestion in section 4(d) is recalled, to examine the distributions in Figs.17(c), (d) and (e) in (II), which were obtained in a spark chamber experiment to observe the $p_T$ distribution from some $10^8$ decays of $K^0_{e3}$. The ordinate in those diagrams is the percentage departure of the observed bin content from a Monte Carlo calculation of the expected (smooth) distribution: the data in the original bins of 0.003 GeV/c in $p_T$ has been converted here to bins of 0.005 GeV in $M_{\mu\pi}$ to make Fig.8(d). As in the original diagram, it has a structure which resembles in form and location the structures in Figs.8(a),(b) and (c). With the large numbers of combinations in the distribution ($\sim 10^7$) it is unlikely that a structure of this amplitude is due to random fluctuations. A postulate that $M_{\mu\pi}^*$ occurs in $K^0_{e3}$ decays violates lepton conservation.

The widths of the "enhancement and lobes" cannot be expected to be as narrow in Fig.8(d) as in the usual $M_{\mu\pi}$ distributions because, presumably, $p_\nu$ may have any laboratory orientation, giving values of $0<p_T<p_\nu$. If the decay which produces $M_{\mu\pi}^*$ were isotropic with respect to the line of flight of the parent kaon, then $p_T \sim p_{Tmax}$ (~0.064GeV/c) would preferred. That any structure is visible accords with the earlier observations of a preferred orientation of the ($\mu,\pi$) combinations with respect to their line of flight, so that $p_T \sim p_{Tmax}$, in contrast to $p_T \sim 0$ in Fig.8(e). No explanation has been found for the apparent lobes in Fig.8(d): it is to be noted, however, that as with e.g., Fig.8(c), the lobes occur together with the "0.429" enhancement, while in Fig.8(e) the 3 features are absent together, according with a postulate that they are associated.

If another mode of decay of $M_{\mu\pi}^*$, as postulated in (II), were responsible for these lobes, then it might be inferred from these observations, that the direction of the neutrino from such a decay might be related to the direction of the neutrino from the parent $K^0_{\mu3}$ decay. Similarly, if the experimental evidence of the $M_{\mu\pi}^*$ enhancement and the lobes being:

(i) associated with $p_T \sim p_{Tmax}$, as in Fig.8(d),
(ii) absent in data with $p_T=0$, as in Fig.8(e), and
(iii) more evident in combinations with unequal laboratory momenta,

is considered together, then it might be inferred that the momenta of the various entities involved in the production and decay of $M_{\mu\pi}^*$ have a tendency for some degree of coplanarity with respect to the line of flight of the parent kaon.

Evidently, only "electronic" detectors can yield the numbers of observations in the distributions in Figs.8(a),(b),(d), and (e). The possibilities to observe any of the structures in those diagrams were not part of the initial aims of the experiments, they were due to publications from the groups and some subsequent kind assistance.

(7) **Averaged structures**

The numbers of values in the distributions in the range $0.380<M_{\mu\pi}<0.470$ GeV, which yield the structures already shown, vary from 14 in Fig.5(c) to some $10^7$ in Fig.8(d). Any features in a distribution obtained by the direct addition of the bins of all the $M_{\mu\pi}$ values in the original data sets would not be a meaningful representation of the information contributed by each experiment. A more equitably weighted representation is in Fig.9, which shows the results of averaging the amplitudes[19] in each interval of 0.0025 GeV in the structures of groups of the various data sets.

Fig.9(a) shows the averages of the amplitudes in the structures from the kaon experiments which yielded Figs. 2(a),(b),(c) and Figs. 8(a),(b),(d).

Fig.9(b) shows the averages of the amplitudes fom the neutrino experiments which yielded Figs. 1(a),(b), Figs. 5(b),(c) and Figs. 7(b),(d).

Fig.9(c) shows the averages of the amplitudes of all the structures in (a) and (b).

Logically, the latter is the most representative structure. However, either of its contributors (a) and (b), which together show compatible features, (with the limitations of the coarser binning and mass resolution of different neutrino experiments), could be the most useful for comparison with the results of an experiment of the same type.

Since the averaged amplitudes for sufficiently large groups of structures composed of random fluctuations will tend to a value of 1.0 in this type of diagram, it should be unnecessary to draw attention to the improbability of some 12 random structures producing the results in Fig.9.

With the aim of assembling what seems appropriate information, it is recalled that the enhancement in the $M_{\mu\gamma}$ distributions in the range $0.422 < M_{\mu\gamma} < 0.437$ GeV, considered as an indication of the possible existence of a charged heavy lepton[20] with a decay $M_{\mu\gamma}^* \to \mu + \gamma$, also occurs in the region of the 0.429 GeV enhancement in Fig.9.

(8) **Conclusion**

(a) **What has been observed?** Structure in distributions in the range $0.380 < M_{\mu\pi} < 0.470$ GeV has been reviewed in observations from 7 independent experiments with kaon decays and from 4 independent neutrino experiments (7 if antineutrino experiments are considered separately). No artifact has been found to explain any part of a structure in the selected range. The mutual compatibility of these structures shows, beyond any reasonable doubt, that they are not recurring random fluctuations.

The only tenable explanation is that the structure is of an unrecognised physical origin. It is also clear that no one data set, alone, can demonstrate beyond any doubt, that its particular structure is of physical origin. The certainty of a physical cause derives from the mutual compatibility of the structures. Their principal features are:

**An enhancement at 0.429±0.002GeV.** The evidence of an enhancement at "0.429 GeV" accords with an hypothesis that it is due to $(\mu,\pi)$ combinations from the decay of a narrow $(\mu,\pi)$ state: $M_{\mu\pi}^*(0.429) \to \mu^- + \pi^+$ and, with reversed charges, similarly for the antiparticle. A lifetime range of $10^{-21}$ to $10^{-12}$ sec is estimated in (II).

**Depletions adjacent to the enhancement.** In most of the $M_{\mu\pi}$ distributions in the vicinity of the enhancement in the "0.429 bin" there are regions where the level is below the level of a fitted mean. The different type of depletions in Fig.3(a) can be attributed to the combinations from the $M_{\mu\pi}^*(0.429)$ decays being unable to fulfil the kinematical constraints for detection in the experiment because their line of flight did not coincide with that of the parent kaon beam. The causes of the usually observed depletions between the enhancement and the lobes are unknown. As with the other features, it is unlikely that the depletions are due to random fluctuations.

**Lobes in the vicinity of 0.405 and 0.445 GeV.** In all except two of the structures there is a lobe on each side of the "0.429" enhancement. Unlike the constancy of location of the enhancement, the lobes are centred within a mass range of some 0.010 GeV about their mean positions, as shown in Fig. 9(c). Because of their slightly varying locations it is unlikely that they are caused by other narrow $(\mu,\pi)$ states: the lobes are as unlikely as the 0.429 GeV enhancement to be due to random fluctuations.

The observations that the places where the lobes and enhancement usually appear are occupied by depletions in Fig.8(e), and are seen in Fig.8(d), accord with lobes being related to $M_{\mu\pi}^*(0.429)$.

There are no corresponding lobes in Figs.1(a) and (c). It is unlikely that their respective $(\mu,\pi)$ combinations would be contaminated with $(\mu,\mu)$ combinations. An hypothesis that the lobes are caused by $(\mu,\mu)$ combinations from another mode of decay of $M_{\mu\pi}^*(0.429)$ is in (II).

**Where do these features occur?** The $M_{\mu\pi}^*(0.429)$ enhancement can be found in an $M_{\mu\pi}$ distribution from neutrino event energies averaging some 90 GeV. The lobes and depletions can be found in an $M_{\mu\pi}$ distribution from neutrino event energies averaging some 62 GeV. Thus, the structure at and around "0.429 GeV" has been found not only in kaon decays, but also in neutrino interactions at widely different average energies.

(b) **What is known?**

My response is that the hypothesis that the "0.429 GeV" enhancement is due to the decay of a short-lived narrow $(\mu,\pi)$ state, accords with all the observations of that feature, and is presumed sound. As well as a decay mode of $M_{\mu\pi}^*(0.429)$, there is a phenomenological upper limit to its lifetime of $10^{-12}$ sec.

As important as this enhancement has been in these studies, it is only one feature in the structure found with it. Apart from the recurrence of the structures indicating the certainty of their physical origin, nothing else could be presumed to be known. A summary of some of the obvious questions about those other features, ordered according to their proximity to the "0.429 GeV" enhancement, is:

(i) Why are there depletions near the enhancement?
(ii) What are the combinations which cause the lobes?
(iii) How sound is the hypothesis of another mode of decay?
(iv) Does the presently observed structure continue beyond the lobes?
(v) Are there other regions with structures in more extensive $M_{\mu\pi}$ distributions?
(vi) Are the indications of the possible existence of a charged heavy lepton $M_{\mu\gamma}^*$, with a mass similar to $M_{\mu\pi}^*(0.429)$, related to these structures?
(vii) Is the physical cause of some, or all, of the features of the structures as yet unidentified artifacts?

There is no final answer to (vii): my own response is that while being well aware of that possibility for some three decades, I still do not know of an artifact in the range of $0.380 < M_{\mu\pi} < 470$ GeV.

(c) **What needs to be done experimentally?**

The long standing aim to determine the identity and momentum of every particle from $K^0_{\mu 3}$ decays, without restriction of their laboratory momenta, is still the most important experimental requirement. Obviously, attempts to meet this need have gone to the limit of feasibility in many experiments - but the progress is insufficient to resolve the questions raised by the existence of structures. Clearly:

(i) The disparate momenta in $(\mu,\pi)$ combinations in the "0.429" enhancement

accentuate the need to detect combinations over their entire ranges of momenta.
(ii) The possibility that the observed lobes may not be due to (μ,π) combinations, accentuates the need to identify all combinations.
(iii) Will an understanding of the depletions require detection of other particles?

If means are invented for obtaining adequately numerous data from new neutrino experiments, their needs will be the same.

(d) **What else needs to be done?**

As stated already, the totality of these compatible experimental results in the range 0.380<$M_{\mu\pi}$<470 GeV is overwhelming evidence that no accepted physical phenomenom is able to account for any of the structure. Therefore it must be assumed that the origin of the structures involves new physics.

In such a situation the credibility of the scientific method of modern particle physics is eroded if evidence of inexplicable phenomena, such as the structure in the range 0.380<$M_{\mu\pi}$<470 GeV, is dismissed without any scientific investigation, with an untenable hypothesis of recurrent statistical fluctuations.

There is no means of excluding that some of the observations in that range may yet have a trivial explanation. Until that is found, Einstein's dictum might offer guidance:

"Raffiniert ist der Herr Gott, aber boshaft ist er nicht".

**Acknowledgments**

Except for one data set[17], the information in this note is, essentially, contained in (II). The present text is aimed to enable and encourage more direct and detailed examination of structures.

My indebtedness to my late wife and the others whose help was very important to me, as mentioned in earlier studies, and to those acknowledged in (II), increases with time. I acknowledge each one of them again, warmly, with sincere appreciation and much nostalgia.

**Notes and references**
1. C.A.Ramm, Phys. Rev.D 32,123 (1985).
2. C.A.Ramm, Phys. Rev.D 26,27 (1982).
3. Mostly a fourth degree polynomial is adequate for the diagrams here. (see note 8).
4. So that the diagrams will simulate the upper part of the original $M_{\mu\pi}$ histograms, without any negative regions, an arbitrary 1.0 (to set the level at which the mean passes through the diagram) is added to (y-m)/(emax-emin): y is the original bin height, m the value of the mean in that bin, emax and emin the extreme values of (y-ym) over the range of $M_{\mu\pi}$ in the diagram. The markings at intervals of 0.0025 GeV are to facilitate comparisons between diagrams with different bin widths.
5. The reasons for confining these studies to the chosen range of $M_{\mu\pi}$ were related to the diversion of effort required to establish an adequate map and treatment of artifacts beyond that range: not to any knowledge of the properties of $M_{\mu\pi}$ distributions.
6. In this report there is the usual acceptance that all observations are subject to random fluctuations. Conclusions take account of estimates of the probability of occurrence and recurrence of the fluctuations. Thus it is assumed that any (or all) of the enhancements described as at "0.429 GeV" could be statistical fluctuations in the $M_{\mu\pi}$ distributions. A conventional estimate is that the probability is vanishingly small ($<10^{-10}$) that the enhancements at 0.429 GeV in (II) are caused by a concurrence of upward random fluctuations at that particular mass. For some, this author included, the occurrence of similar structures in independent distributions tends to negate an hypothesis that those structures are random fluctuations.

7. In order to minimise any error as to their nature, the data in the "original" enhancement resulted from a request to the scanners to select only $(\mu,\pi)$ combinations which, by phenomenology, were categorised as certain to contain a pion.

8. This experiment (ref. 6 in (II)) was designed to study the $(\pi,\pi)$ decays of $K^0_L$. The combinations were selected in which the $(\pi,\pi)$ had the same laboratory momenta and no resultant momentum transverse to the $K^0_L$ beam direction ($p_T=0$). The absolute mass error in the observed kaon mass was <0.002 GeV. By changing the selected momentum, the same installation could be used to record combinations with $p_T=0$ which, if assumed to include $(\mu,\pi)$, would give a distribution in the range $0.370 <M_{\mu\pi}<0.500$ GeV. After conversion of the observed bins of 0.002 GeV to bins of 0.0025 GeV, the distribution rises from zero to some 3000 entries per bin near 0.44 Gev and then falls to zero. Polynomial means of degrees 10 to 13 were fitted and then averaged to reduce effects from any peculiarities of the individual means. Fig.4(d) in (II) shows the departures of the bin contents from the averaged mean.

It might be asked how likely the positioning of the depletions in Fig.3(a) is due to statistical fluctuations about the mean, such that the largest three downward fluctuations occur at the expected locations of the usual enhancement and lobes? In the 36 bins in the diagram there are 7140 possible arrangements of any chosen three fluctuations. After noting that each of the largest three downward fluctuations has another downward fluctuation adjacent, it can be estimated that the probability of the structure in Fig.3(a) appearing in the locations corresponding to the enhancement and lobes in the other kaon observations is $<2\times10^{-4}$.

9. H.C.Ballagh and 36 co-authors, Phys. Rev.D 21,569 (1980).

10. H.C.Ballagh and 26 co-authors, Phys. Rev.D 29,1300 (1984).

11. For the $(\mu^-,\pi^+)$ distribution the two bins in the range $0.420 <M_{\mu\pi}<0.440$ GeV have been removed before the fitting.

12. It may be useful to some readers to remark that the enhancement in the bin with its lower edge at 0.490 GeV in (e) is in the region where an artifact from $K^0_{\pi\pi}$ can occur (see A2, and Fig.2 in (II)).

13. Of importance to plans for future experiments is the similarity of the number of events in the region of the "0.429" bins from the neutrino experiments at such differing beam energies. As stated in note (7) the 37 combinations of the "original" distribution were selected individually by the scanners; the selection rate of the combinations of Ballagh et al., was about one per 4000 pictures.

14. The yield of 71 values in the range $0.380<M_{\mu\pi}< 0.470$ GeV added some new information to these studies. It also makes clear that the experimental study of the composition and angular distributions of the enhancement and the neighbouring features requires a much more prolific data source than might be expected from a high energy bubble chamber neutrino experiment.

15. As described in Table 1, AE might be considered as a very approximate indicator of signal to background for departures of a histogram from its mean. Evidently, if the bin levels follow the mean exactly, then AE=0. Rebinning the data in larger bins would usually diminish AE, as would also a greater background continuum. A low value of AE in Table 1 is for the $M_{\mu\pi}$ distribution from the HBC where the "(μ,π)" included the charged combinations frrom $K^0_{\pi\pi}$, $K^0_{\mu3}$ and $K^0_{e3}$. Nevertheless Fig1.(c) is an informative diagram. The greatest value of AE is for the data of Fig.1(a) where the bin width and location were chosen to demonstrate the congregation of the $M_{\mu\pi}$ values.

16. D.Allasia and 47 co-authors, Nucl. Phys. B224,1 (1983).
17. D.Allasia and 36 co-authors, Phys. Rev.D 31,2996 (1985).
18. Some readers may notice that the possible kaon artifacts referred to in connection with Figs. 4(a) and 6(d) are actually one bin apart. If these features are, in fact, attributable to that artifact, their different positions could imply that the mass scales of Ballagh et al. and Allasia et al. do not correspond exactly in the region of 0.490 Gev. In order to bring the artifacts into line the structures of Allasia et al. in Figs. 7(b) and (d) would need to be displaced one bin to the right. Such a displacement would enhance the alignment of all the structures in Fig.7. Evidently, a reverse displacement would apply for Ballagh et al., however the observation of the kaon mass shown by those authors confirms the calibration of their mass scale in that region. In the latter regard, it has been noticed that the width of the $M_{\mu\pi}^*$ enhancement is sometimes narrower than might be expected from observation of the width of $K_{\pi\pi}$ feature. This could be due, in part, to both tracks for $K_{\pi\pi}$ being subject to nuclear scattering, as distinct from only one for $M_{\mu\pi}^*$.
19. Note 4, which concerns most of the preceding figures, explains how the level of the mean for each structure is set at 1.0, and the scale distance between the extreme amplitudes is set at one unit on that scale. Averaging the amplitudes in a given mass interval for a group of such structures produces an "averaged structure" in which each data set has the same weight. Evidently, the level of the means will average to 1.0. Unlike the extreme range of the amplitudes in the original structures being constrained to a constant value of 1.0, the extreme (but unlikely) range for averaged structure is 0<(averaged amplitude)<2.0.
20. C.A.Ramm, Nature 230, 145 (1971).

**(A1):The inability to distinguish muons from pions.** Some effects due to a lack of muon-pion identification may be illustrated by supposing a source of neutral (μ,π) combinations of various laboratory energies, all with the same invariant mass $M_{\mu\pi}$ and with no preferred orientation of the centre of mass system (cms) to their laboratory line of flight. If there is no means of particle identification, what $M_{\mu\pi}$ distribution can be observed?

If the invariant masses are calculated for each possible assignment of the combinations, half the values will be $M_{\mu\pi}$, the others (reverse assigned) will each have an $R_{\mu\pi}$ in a range extending above and below $M_{\mu\pi}$, depending on their laboratory energies and the actual value of $M_{\mu\pi}$. With no preferred orientation in the cms, the numbers of the $R_{\mu\pi}$ will increase nearer to $M_{\mu\pi}$: without particle assigment the observed $M_{\mu\pi}$ distribution is not a single line at the constant invariant mass from the source. Of course in this example, with a single known value of $M_{\mu\pi}$, the assignment can be deduced, except for combinations of μ and π of equal momenta.

Suppose again the same source but which now gives a configuration in the cms of the (μ,π) so that the preferred direction of the actual muon is always in the forward direction with respect to the laboratory line of flight. From the correct assignments the line at $M_{\mu\pi}$ will appear as before, accompanied by a grouping of the $R_{\mu\pi}$ values at a higher mass than $M_{\mu\pi}$. The actual centre and extent of the spread of the $R_{\mu\pi}$ will depend on the energy distribution of the source. (Fig.5 in (I) shows the possibilities).

**(A2): Artifacts.** Experimental artifacts which might have caused reproducible enhancements in the various data sets have been sought intensively. None has been found in the range $0.380<M_{\mu\pi}<0.470$ GeV. A commonly observed artifact is near 0.49GeV, from the $(\mu,\pi)$ assignment of $(\pi,\pi)$ from the $K^0_{2\pi}$ decay. The next closest artifact from kaons is in the vicinity of 0.36 Gev, from $(\mu,\pi)$ assigment of the $(\pi,\pi)$ with the highest invariant masses from $K^0_{3\pi}$. Both of these artifacts are mentioned in (II).

**(A3): Calibration.** Mass scales were verified by observations of $K^0_s \rightarrow (\pi^+,\pi^-)$. In some experiments great stability in invariant mass scales was not important for the achievement of the original aim. In one such case, with detailed guidance from the experimenters, it was possible to correct for most long term shifts in the mass scale ((II) Fig.3). Evidently, random errors in mass measurements are incorrigible.

**(A4):Initial search procedures.** For $(\mu,\pi)$ in the $M_{\mu\pi}^*(0.429)$ enhancement the preferred direction of the muon is in the forward hemisphere with respect to the line of flight. Thus, if the $(\mu,\pi)$ combinations have laboratory momenta $p_1$ and $p_2$, such that $p_1>p_2$ and the particles are assigned independently of any identification, then the invariant mass distribution from the assignment of $p_\mu = p_1$ will tend to have a more pronounced enhancement than the distribution from $p_\mu = p_2$. Constraining the transverse momentum ($p_T$) of the combinations, with respect to the line of flight of the kaons, to $p_T<0.100$ GeV/c can be useful for reducing a background of combinations from $K^0_{e3}$. For the combinations from $M_{\mu\pi}^*(0.429)$ $p_{Tmax}$ is 0.064 GeV/c.

**Table 1**

| Data | No. | nb | bw | u | d | m | AE | a |
|---|---|---|---|---|---|---|---|---|
| 1nl | 37 | 5 | 0.020 | 11.8 | 2.5 | 6.2 | 0.62 | μlc, πs |
| 2nlg | 368 | 18 | 0.005 | 11.3 | 5.8 | 20.3 | 0.23 | μlc |
| 3kh | 2334 | 36 | 0.0025 | 19.9 | 16.1 | 74.1 | 0.09 | none |
| 4kl | 908 | 36 | 0.0025 | 15.6 | 14.7 | 24.4 | 0.17 | eu |
| 5k2m | 3784 | 18 | 0.005 | 32.4 | 27.7 | 200 | 0.06 | none |
| 6kc | 62020 | 36 | 0.0025 | 64 | 64 | 2082 | 0.008 | none |
| 7f- | 57 | 9 | 0.010 | 7.1 | 3.8 | 6.87 | 0.37 | EMI |
| 8f+ | 14 | 9 | 0.010 | 1.3 | 1.4 | 1.60 | 0.20 | EMI |
| 9f-+ | 71 | 9 | 0.010 | 7.5 | 5.1 | 8.47 | 0.37 | EMI |
| 10c- | 232 | 9 | 0.010 | 7.7 | 10.3 | 26.9 | 0.20 | none |
| 11c+ | 143 | 9 | 0.010 | 5.2 | 6.1 | 15.8 | 0.24 | none |
| 12c-+ | 375 | 9 | 0.010 | 14.0 | 10.5 | 42.6 | 0.15 | none |
| 13ki | 11310 | 36 | 0.0025 | 75 | 38 | 332 | 0.05 | none |
| 14kb | 114900 | 36 | 0.0025 | 161 | 115 | 4489 | 0.013 | none |
| 15kw | >10$^7$ | 18 | 0.005 | 0.61* | 0.35* | 100* | 0.003 | none |

\* excursions are percentage departures from the expected bin content

**Information for Table 1 in the range $0.380 < M_{\mu\pi} < 0.470$ GeV.**

Columns:
no.      number of entries
nb      number of bins
bw     bin width in GeV
m       level of the mean in the "0.429 bin"
u       maximum upward excursion from the mean
d       maximum downward excursion from the mean
AE    ($\Sigma$ absolute excursions from mean)/(nb x m)
A       assignment of $(\mu,\pi)$: μ by lepton conservation (μlc), μ by EMI (EMI),
          π by scanning (πs), electron contamination unlikely (eu), no identification
          of e, μ, π, p (none)

Rows:    (Data from the referenced figure.)
1nl       Neutrino experiment in HLBC (Nature 227,1323 (1970))
2nlg     Neutrino experiments in HLBC + Gargamelle, (II) Fig.5(a)
3kh      Kaon experiment in HBC, both values of $M_{\mu\pi}$, (II) Fig.2(d)
4kl       Kaon experiment (X4) in HLBC, both values of $M_{\mu\pi}$, (II) Fig.1(a)
6k2m    Kaon experiment 2M HBC, both values of $M_{\mu\pi}$, (II) Fig.6(d)
6kc      Kaon experiment Clarke et al. (II) Fig.4(d)
7f-       $M_{\mu\pi}$ distribution as indicated by $(\mu^-, \pi^+)$ in Ballagh et al.[10] Fig.2
8f+      $M_{\mu\pi}$ distribution as indicated by $(\mu^+, \pi^-)$ in Ballagh et al.[10] Fig.2
9f-+     $M_{\mu\pi}$ distribution from sum of f- and f+
10c-     $M_{\mu\pi}$ distribution as indicated by $(\mu^-, \pi^+)$ in Allasia et al.[16] Fig.1
11c+     $M_{\mu\pi}$ distribution as indicated by $(\mu^+, \pi^-)$ in Allasia et al.[16] Fig.1
12c-+   $M_{\mu\pi}$ distribution from sum of c- and c+
13ki     Kaon experiment Chien et al. (II) ref. 7 and Figs.3(b),4(b),6(e)
14kb    Kaon experiment Bisi et al., (II) ref. 9 and Figs.4(c) and 6(g))
15kw   Kaon experiment Webb, (II) refs. 26, 27 and Figs.17(c),(d) and (e))

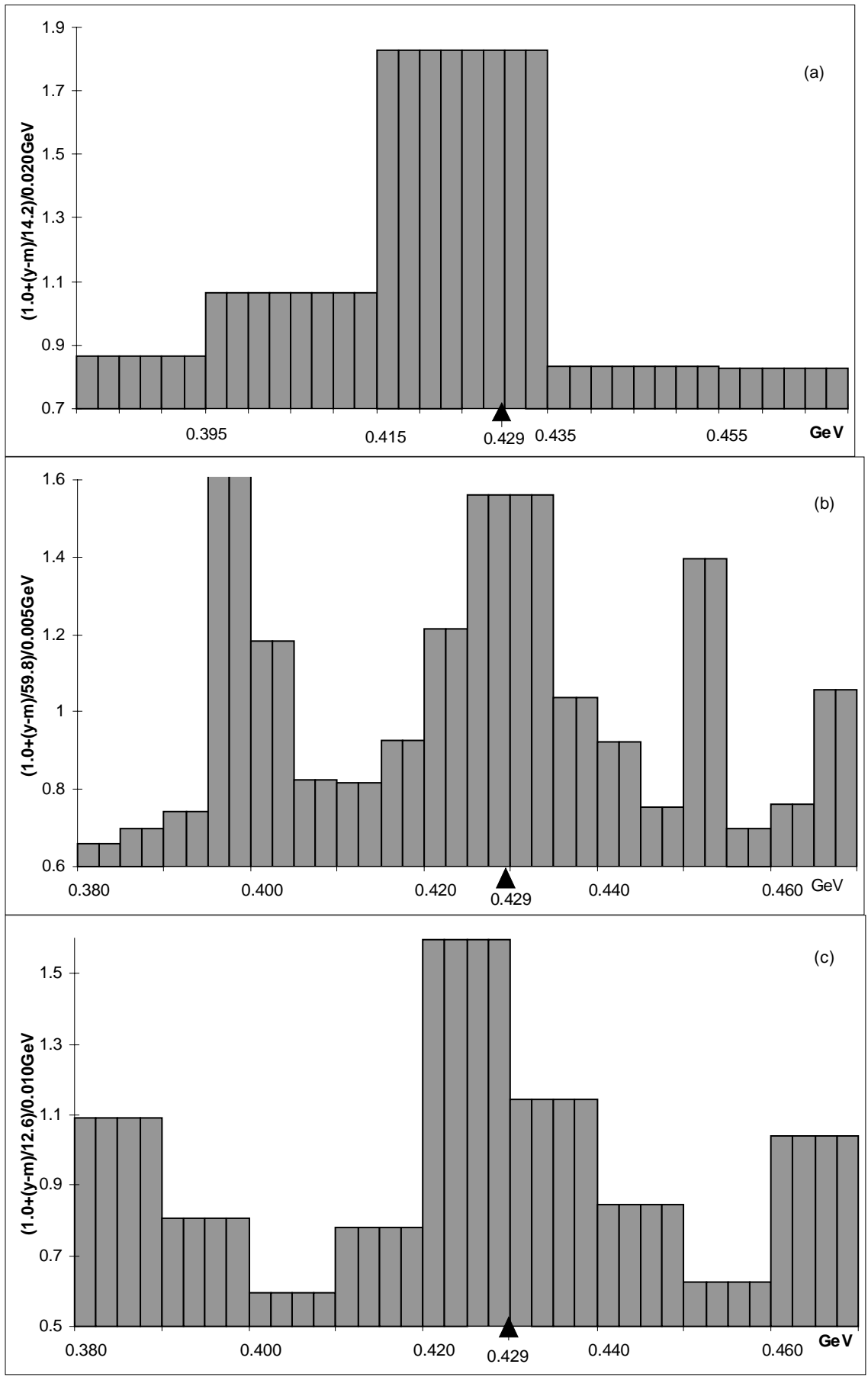

Fig.1. Structure in $M_{\mu\pi}$ distributions from neutrino interactions in:-
(a) The first neutrino experiment in the HLBC, pion track phenomenologically identifiable. The binning was chosen to show the unexpected accumulation of values (Data 1nl).
(b) Data from the HLBC and Gargamelle. Pion assigned as likely to be a pion. (Data 2nlg).
(c) The combined $(\mu^-,\pi^+)$ and $(\mu^+,\pi^-)$ observations.by Ballagh et al. in the Fermilab 15-ft hydrogen bubble chamber with an external muon identifier (Data 9f-+).

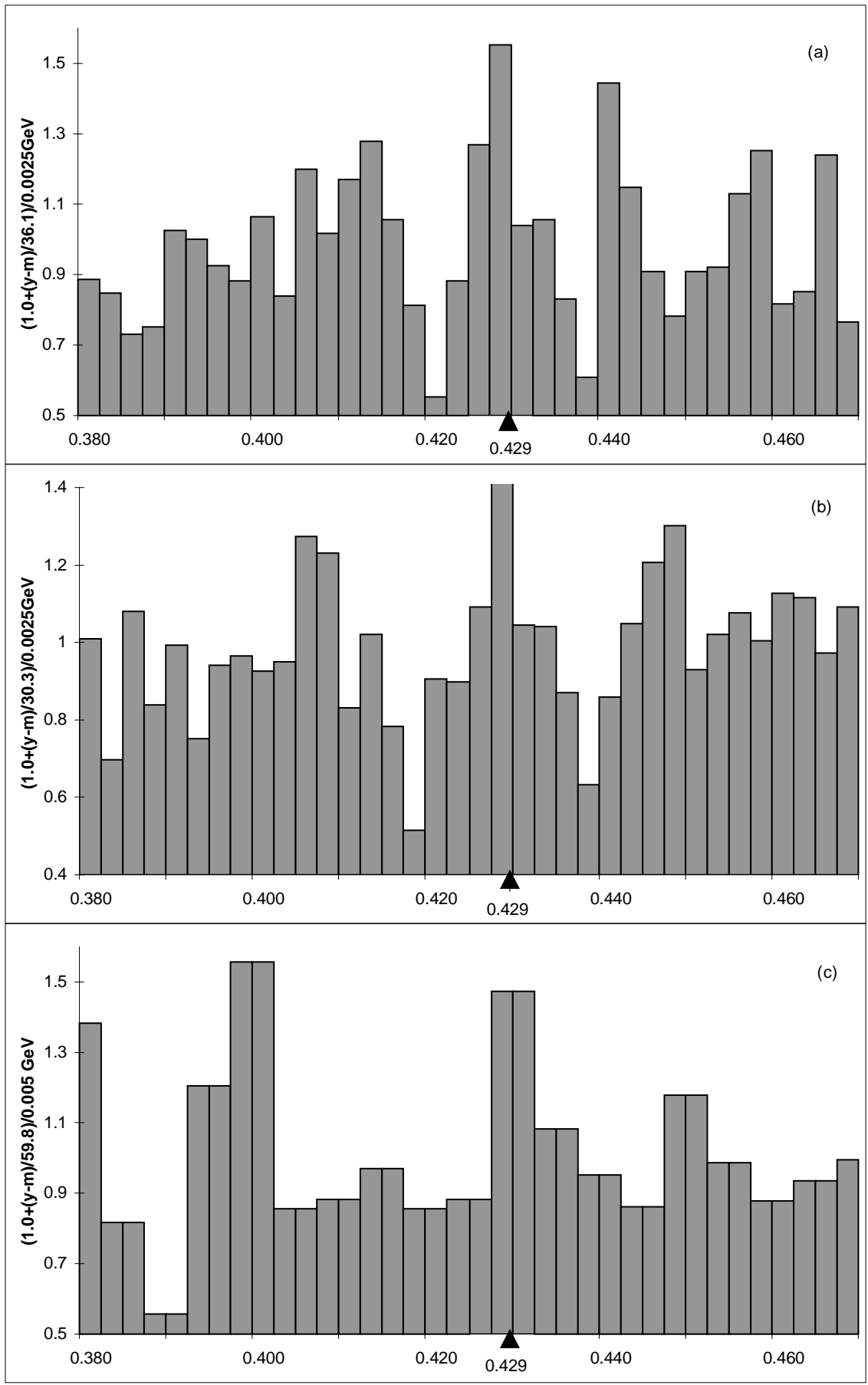

Fig. 2. Structure in $M_{\mu\pi}$ distributions from kaon decays in:-
  (a) An experiment in the HBC. No identification of particles, some background was reduced by the selection $p_T<0.100$ GeV/c (Data 3kh).
  (b) The X4 experiment in the $0.5M^3$ version of the HLBC, fitted with a beam pipe and filled with $CF_3Br$ (radiation length ~11cm) for detection of electrons or $\gamma$-rays (Data 4kl).
  (c) The 2m hydrogen bubble chamber. Tracks mostly unidentifiable (Data 5k2m).

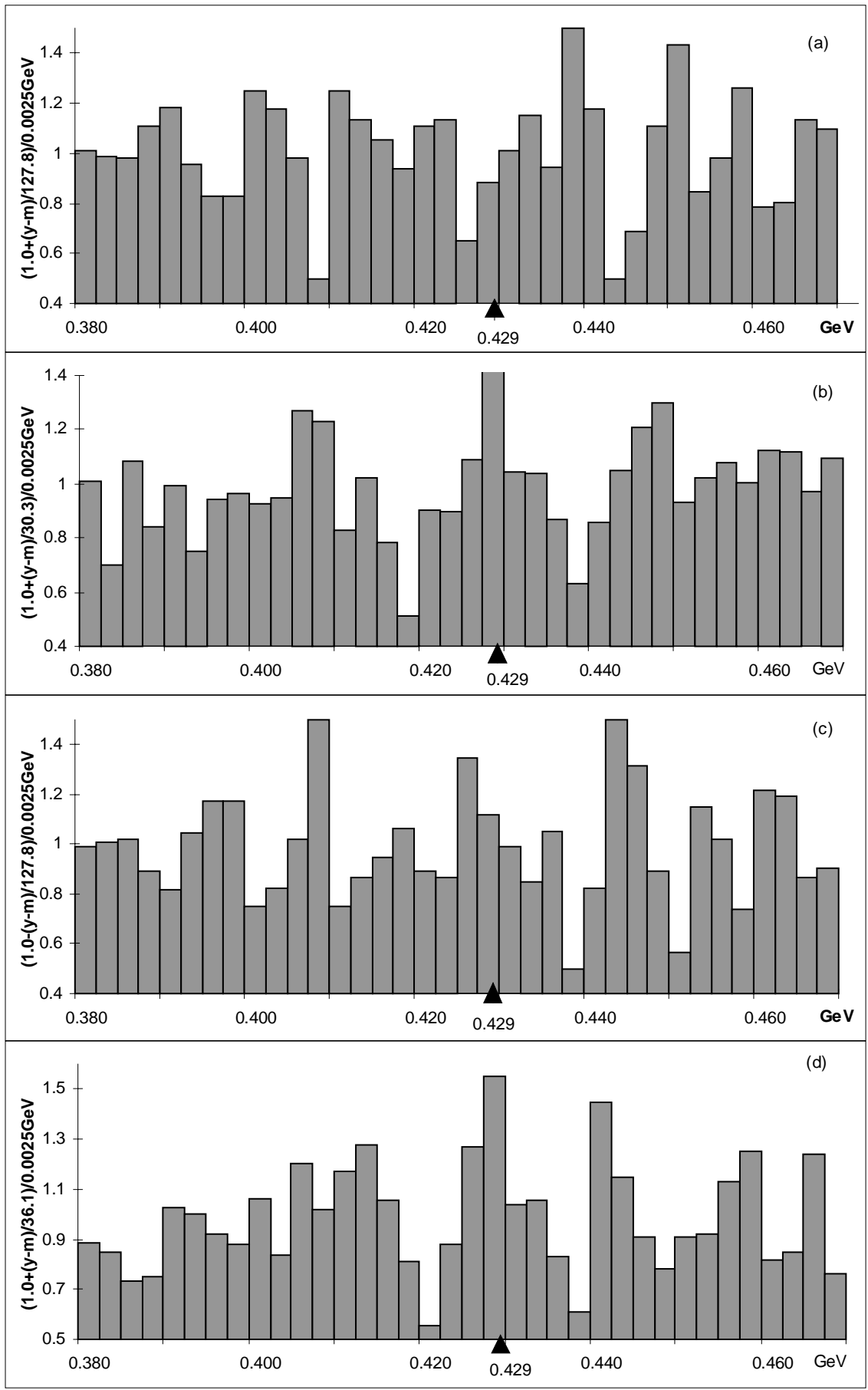

Fig. 3. Effect of $p_T$ in $M_{\mu\pi}$ distributions from kaon decays (see section 3(d) and note 8).
  (a) Structure from $(\mu,\pi)$ selected with $p_T = 0$ (Data 6kc).
  (b) Fig.2(b), unconstrained $p_T$, shown again for comparison. Note the correspondence of the major 3 depletions in (a) with the "0.429 enhancement and lobes".
  (c) The structure in (a) has been inverted to facilitate comparisons (note ordinate scale).
  (d) Fig.2(a), unconstrained $p_T$, shown again for comparison.

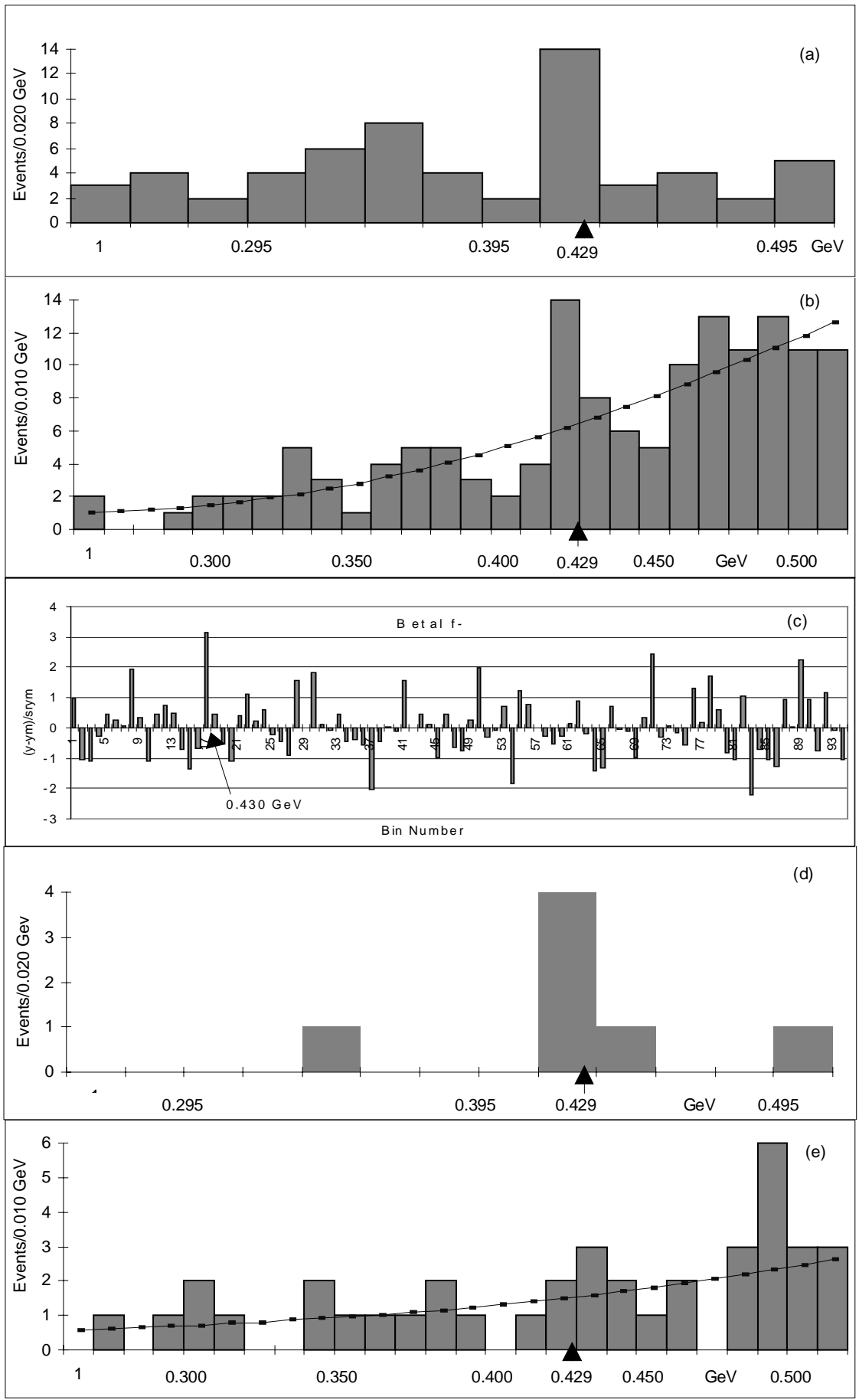

Fig. 4 (a) $M_{\mu\pi}$ values < 0.520 GeV from $(\mu^-,\pi^+)$ from the HLBC, in the data for Fig.1(a).
 (b) $M_{\mu\pi}$ values from $(\mu^-,\pi^+)$ of Fig.2 of Ballagh et al. (Data 7f-).
 (c) Values of $(y-ym)/ym^{1/2}$ (y is the level,, ym the mean) for each of the 94 bins of 0.010 GeV from which (b) is shown.
 (d) $M_{\mu\pi}$ values from $(\mu^+,\pi^-)$ from the HLBC in the data for Fig.1(a).
 (e) $M_{\mu\pi}$ values from $(\mu^+,\pi^-)$ of Fig.2 of Ballagh et al. (Data 8f+, see note 12).

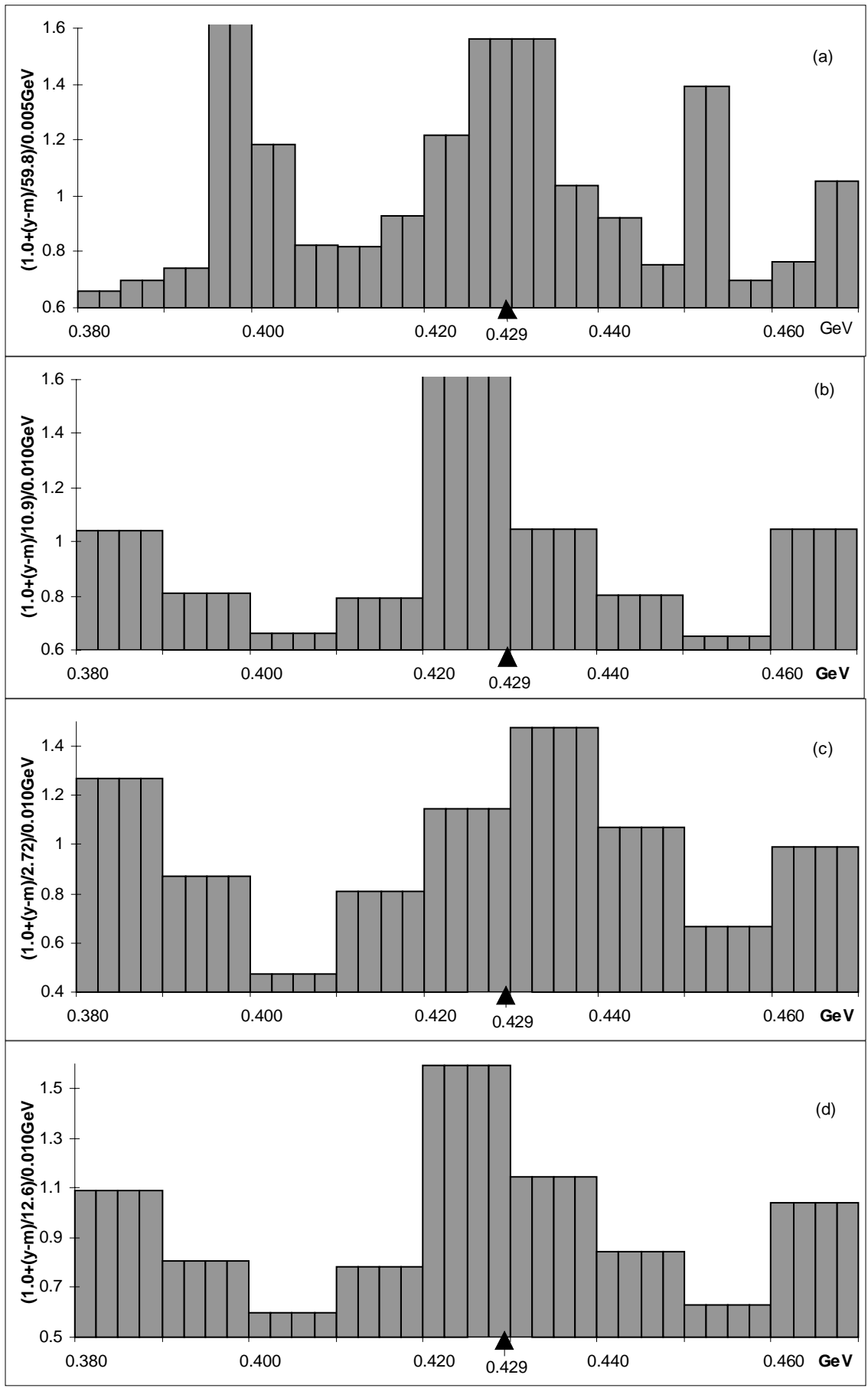

Fig. 5. Comparison of structure in $M_{\mu\pi}$ distributions of Ballagh et al.
   (a) Fig.1(b) from the HLBC and Gargamelle, repeated for reference.
   (b) Structure in the distribution in Fig.4(b), (Data (7f-)).
   (c) Structure in the distribution in Fig.4(e), (Data (8f+)).
   (d) Structure in the combined distributions in Figs.4(b) and (e), (Data (9f-+)).

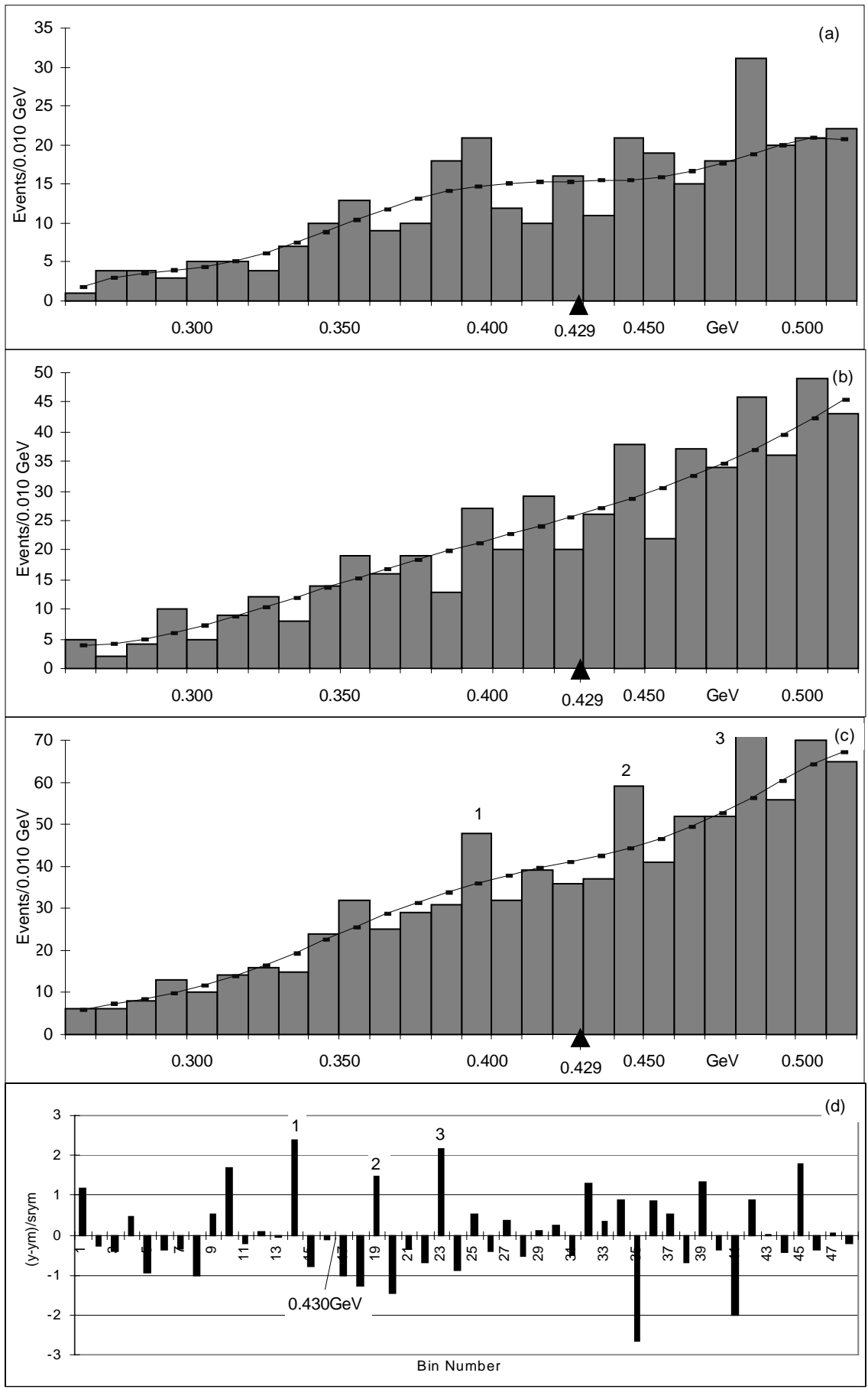

Fig. 6. $M_{\mu\pi}$ values < 0.520 GeV in Fig.1 of Allasia et al.
  (a) $M_{\mu\pi}$ values from $(\mu^+,\pi^-)$. (Data 11c+, see note 12).
  (b) $M_{\mu\pi}$ values from $(\mu^-,\pi^+)$. (Data 10c-).
  (c) $M_{\mu\pi}$ values from the combined data of (a) and (b), (Data 12c-+).
  (d) The values of $(y-y_m)/y_m^{1/2}$ for the 48 bins of 0.010 GeV from which (c) is shown (see caption of Fig.4(c)). Values at 1, 2 and 3 are for the same bins in (c), see section 5(c).

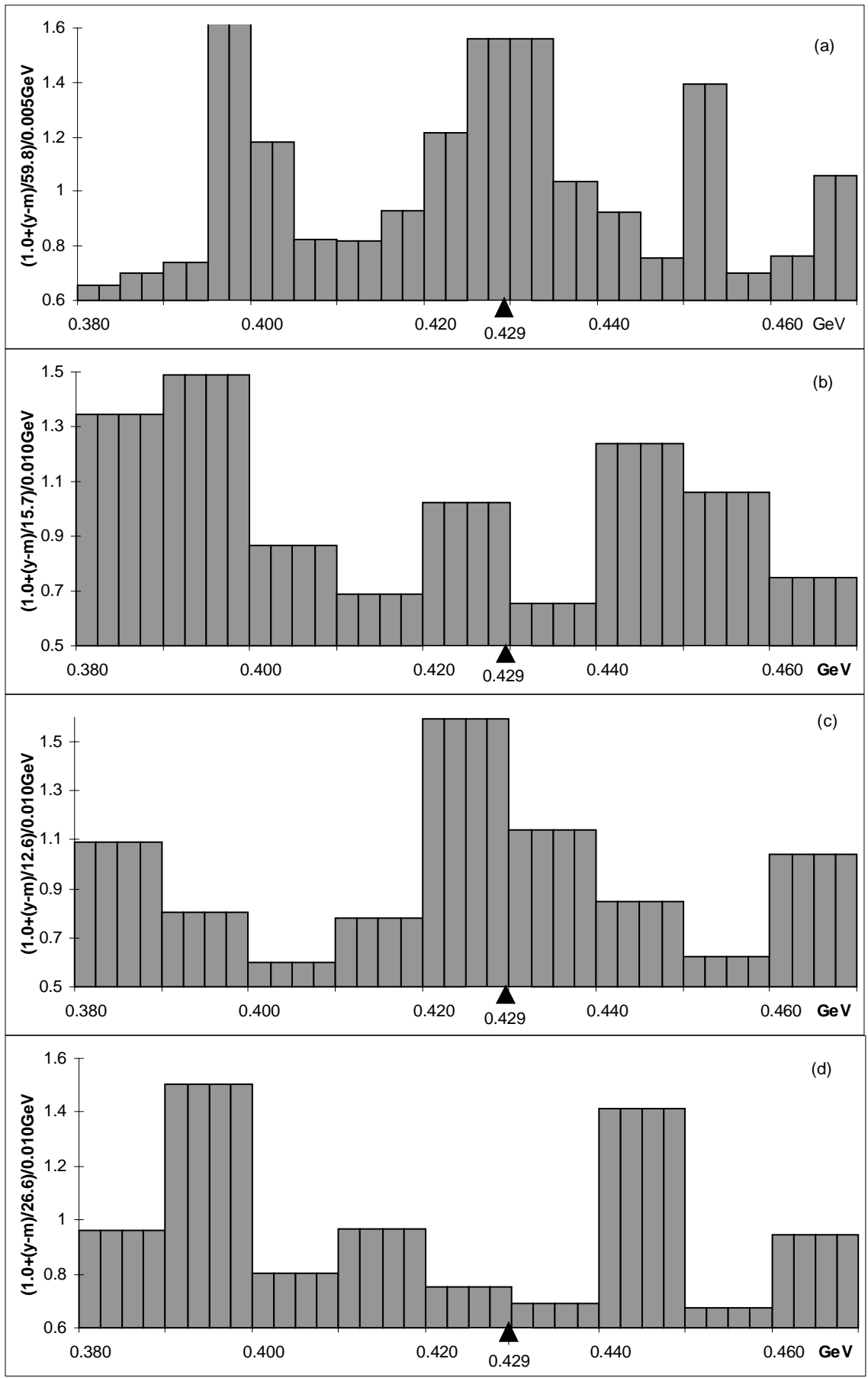

Fig.7. Comparisons of structure in $M_{\mu\pi}$ distributions shown by Allasia et al.
  (a) Fig.1(b) from the HLBC and Gargamelle, repeated for reference.
  (b) Structure in the distribution in Fig.6(a), from $(\mu^+,\pi^-)$, (Data (11c+).
  (c) Structure in the total distribution of Ballagh et al. shown also in Fig.5(d).
  (d) Structure in the distribution in Fig.6(b), from $(\mu^-,\pi^+)$, (Data (10c-). Lobes
     in corresponding places in (a), (b) and (d) are not as prominent in (c).

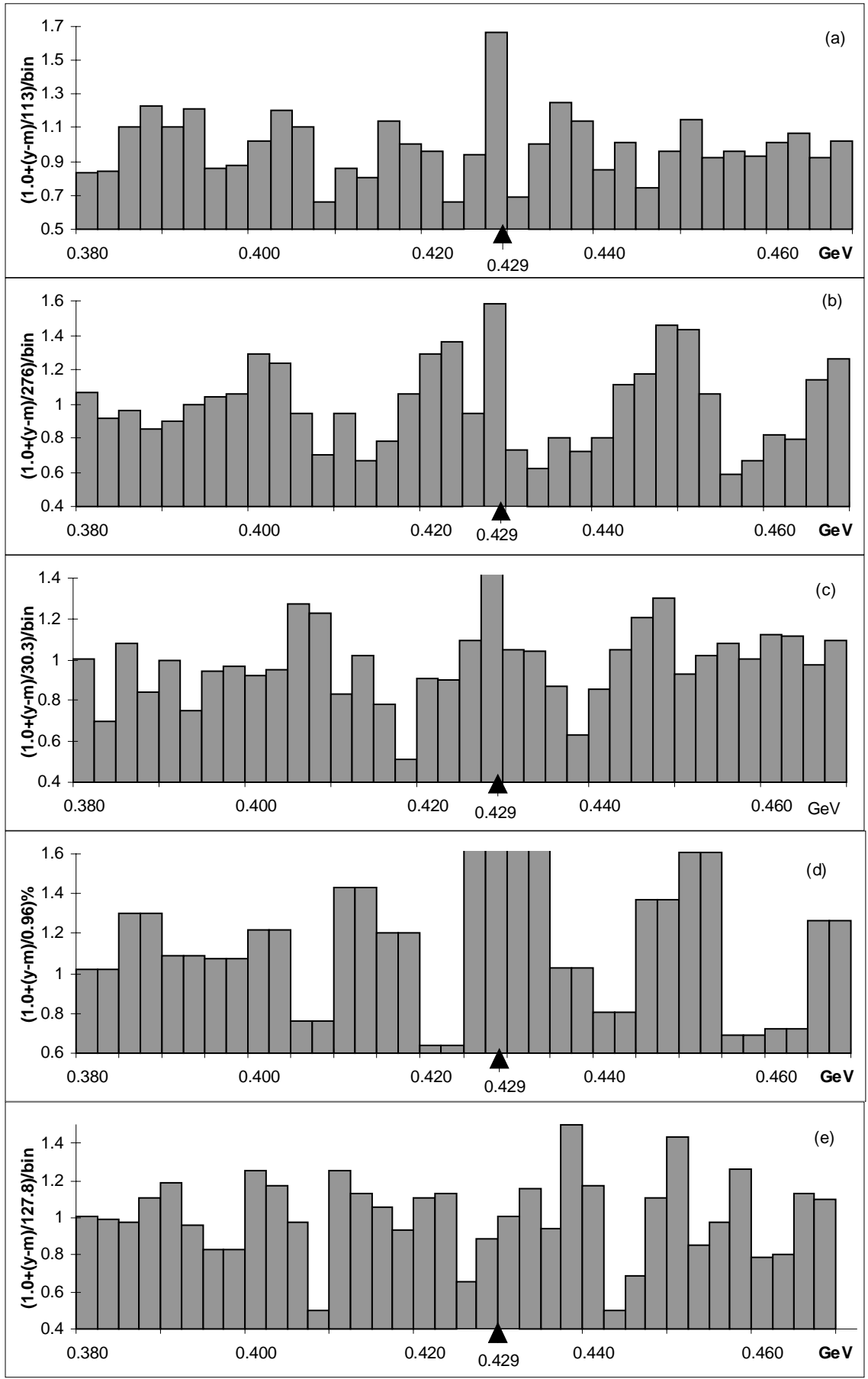

Fig.8. Structures in $M_{\mu\pi}$ distributions from kaon experiments with spark chambers.
 (a) Structure in (II) Fig.3, Fig.4(b) and Fig.6(a), (bins 0.0025 GeV, Data13ki).
 (b) Structure in (II) Fig.4(c) (bins 0.0025 GeV, Data 14kb).
 (c) Fig.2(b) from the HLBC, shown again for comparison (bins 0.0025 GeV).
 (d) Structure in $p_T$, scaled as $M_{\mu\pi}$, from a $K^0_{e3}$ experiment, (bins 0.005GeV, Data 15kw).
 (e) Fig.3(a), shown again to recall the effect of a selection of $p_T\sim 0$, (bins 0.0025 GeV).

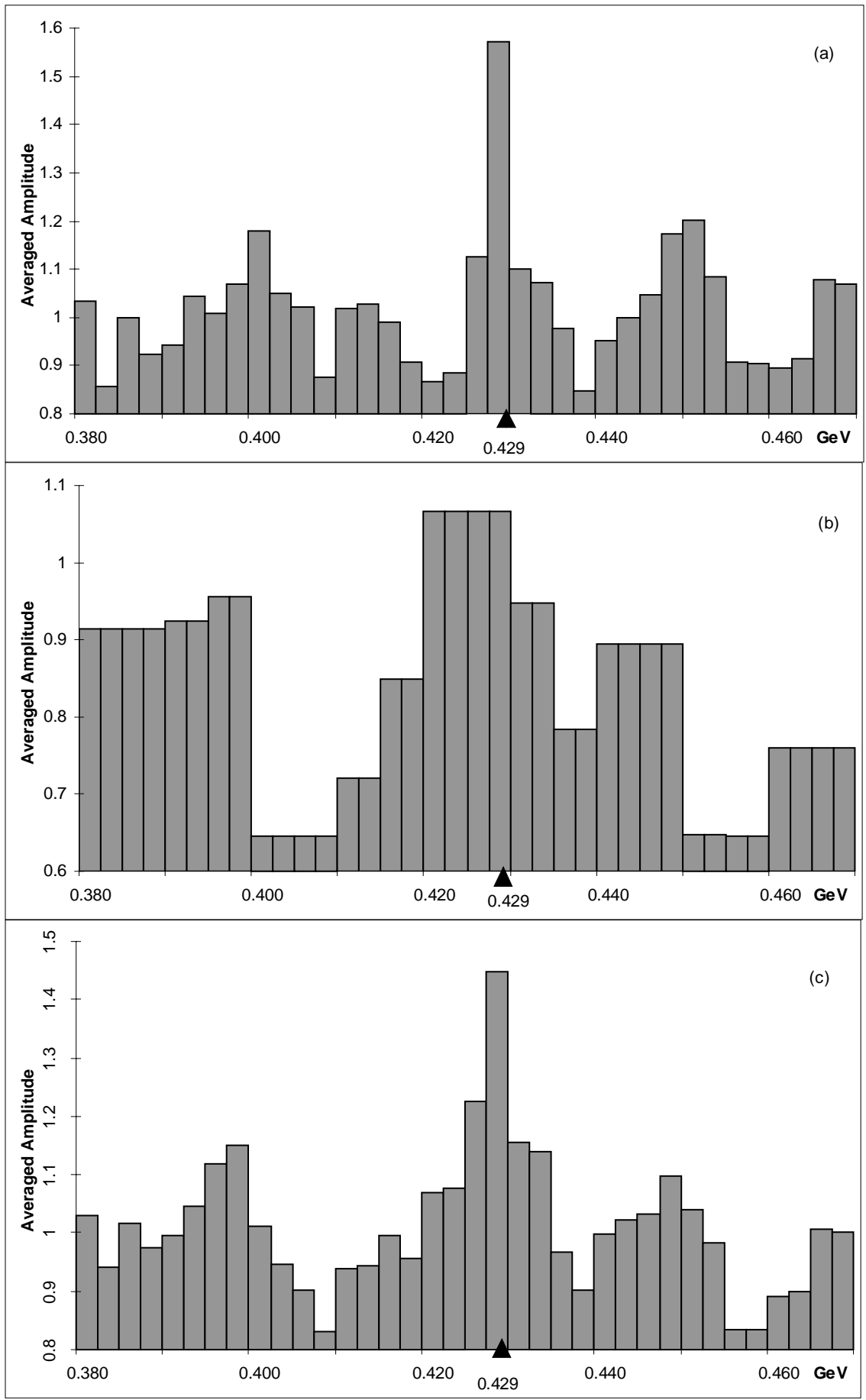

Fig.9. Averaged structures, in intervals of 0.0025 Gev in $M_{\mu\pi}$ distributions from experiments with kaons and neutrinos (see section 7 and note 19).
(a) Averages of the amplitudes in the structures from the kaon experiments which yielded Figs.2(a),(b),(c) and Figs.8(a),(b),(d).
(b) Averages of the amplitudes in the structures from the neutrino experiments which yielded Figs.1(a),(b),(c), Figs.5(b),(c) and Figs.7(b),(d).
(c) Averages of the amplitudes in all the structures in (a) and (b).